# Uchimata: a toolkit for visualization of 3D genome structures on the web and in computational notebooks

David Kouřil[1], Trevor Manz[1], Tereza Clarence[2], Nils Gehlenborg[1,*]
[1] Harvard Medical School, Boston, MA, USA
[2] Icahn School of Medicine at Mount Sinai, New York, NY, USA
[*] Corresponding author: nils@hms.harvard.edu

## Abstract

**Summary:** Uchimata is a toolkit for visualization of 3D structures of genomes. It consists of two packages: a Javascript library facilitating the rendering of 3D models of genomes, and a Python widget for visualization in Jupyter Notebooks. Main features include an expressive way to specify visual encodings, and filtering of 3D genome structures based on genomic semantics and spatial aspects. Uchimata is designed to be highly integratable with biological tooling available in Python.

**Availability and Implementation:** Uchimata is released under the MIT License. The Javascript library is available on NPM, while the widget is available as a Python package hosted on PyPI. The source code for both is available publicly on Github (https://github.com/hms-dbmi/uchimata and https://github.com/hms-dbmi/uchimata-py) and Zenodo: (https://doi.org/10.5281/zenodo.17831959 and https://doi.org/10.5281/zenodo.17832045). The documentation with examples is hosted at https://hms-dbmi.github.io/uchimata/

**Contact:** david_kouril@hms.harvard.edu or nils@hms.harvard.edu.

## Introduction

Alongside sequencing-based technologies developed to probe the spatial conformation of genomes—such as the influential Hi-C method (Lieberman-Aiden et al. 2009)—there are ongoing efforts to produce concrete 3D models that depict genome folding. Although research on macromolecular structures has long benefited from sustained investment in resources such as the Protein Data Bank (Berman et al. 2000) and visualization tools like PyMOL (Schrödinger, LLC 2015) and Mol* (Sehnal et al. 2021), comparable tools for analysis and visualization of physical *genome* structures remain limited.

There are two principal approaches to generating structural models of genomes (Imakaev et al. 2015): *data-driven* methods, which produce structures that satisfy constraints derived from input data (e.g., Hi-C), and methods that simulate structures *de novo* based on mechanistic principles. Li et al. (Li et al. 2023) review reconstruction methods and highlight how resulting 3D structures can lead to novel insights. Portillo-Ledesma et al. (Portillo-Ledesma et al. 2023) further emphasize that the choice of model resolution and simulation technique restricts the scale of the phenomena that can be investigated. Oluwadare et al. (Oluwadare et al. 2019) summarize approaches for reconstructing

chromosome- and genome-level structures from Hi-C data. Notable examples of existing structures include Duan et al.'s yeast genome model (Duan et al. 2010), Stevens et al.'s mammalian (mouse) genome structures constructed from single-cell Hi-C (Stevens et al. 2017), and Tan et al.'s genome structures of single diploid human cells (Tan et al. 2018). Recent studies show promise in employing machine learning techniques to infer 3D genome structures (Schuette et al. 2025). Beyond individual publications, there have been efforts to develop centralized databases of 3D genome models (Oluwadare et al. 2020), although community adoption remains limited.

Visualizing specialized three-dimensional data can be challenging for computational biologists lacking computer graphics expertise, who depend on tools that abstract low-level rendering details. Although several tools have been developed for *visualizing 3D genome* structural data—such as Genome3D (Asbury et al. 2010), 3DGB (Butyaev et al. 2015), GMOL (Nowotny et al. 2016), HiC-3DViewer (Djekidel et al. 2017), and CSynth (Todd et al. 2021)—most exist as standalone desktop or web applications, making them difficult to integrate into notebook-based analytical workflows, to adapt for case-specific applications, and to customize beyond the visual choices predefined by their developers.

Two genome browsers—WashU Epigenome (Li et al. 2022) and Nucleome Browser (Zhu et al. 2022)—currently provide visualization capabilities for 3D genome models. These remain the most comprehensive toolkits available, integrating diverse genomics data types and visualizations. However, their integrative power still falls short compared to computational notebooks, which enable flexible combination of software packages through end-user programming (Ko et al. 2011) to support novel analytical scenarios. Computational notebooks have become the de facto standard for analysis in computational biology and are commonly used to interface with simulation packages that generate 3D genome structures. Despite this, visualization support for such data within notebook environments has lagged behind.

To address the lack of dedicated visualization tools in computational notebooks, scientists often repurpose general plotting libraries or tools originally designed for protein structures. For instance, nglutils (https://github.com/mirnylab/nglutils) builds on the molecular viewer nglview (Nguyen et al. 2018) to support interactive visualization of genome structures within notebooks. Similarly, some simulation packages for 3D genome structures recommend using general visualization libraries, such as matplotlib (Hunter 2007) or fresnel (https://github.com/glotzerlab/fresnel). Biologists often reuse molecular file formats (e.g., PDB) to store 3D genome structures, but differing interpretations of these standards for genomic data can cause incompatibilities even within the same format.

We developed **uchimata**, a toolkit for visualizing 3D genome structures across the web and computational notebook environments (Fig. 1). Its unique combination of features exceeds the capabilities of existing solutions: (1) **A declarative specification for visual encoding** provides flexible and expressive customization of the visual depiction and allows mapping genomic data onto the 3D structure; (2) **An API for spatial and genomics filtering** of 3D genome structures supports operations such as applying cutting planes or selecting genomic ranges. (3) **Dual JavaScript and Python availability** enables use both during analysis—close to the modeling and analysis environment—and as an integrated component within larger web applications, such as data portals.

At its core, uchimata operates on tabular data that link genomic regions to XYZ coordinates, making it suitable for a broad range of applications, including Hi-C-based simulations, de novo modeling, and emerging approaches such as imaging-based methods.

# Methods

## Design of the uchimata toolkit

The design of uchimata is guided by the **principle of composability** (Manz 2024). To support long-term maintainability and enable the reuse of visualization tools as components in novel scenarios, each tool should focus on a narrow and well-defined set of features. This contrasts with the approach of many other tools in this domain, which combine 3D data with Hi-C matrices and genomic browsers, ultimately reducing reusability. In uchimata, our focus is on visualizing 3D models, while related representations are deliberately left to other specialized tools. This modular approach allows users to select and combine components to construct interfaces and pipelines tailored to their specific requirements, avoiding the pitfalls of monolithic software that becomes progressively difficult to extend and maintain.

The second guiding principle is the **separation of concerns** and leveraging strengths of individual programming environments. Over time, the Javascript and Python ecosystems have developed distinct competencies. The web platform excels at building user interfaces and offers a wide range of visualization libraries. Its core technologies—HTML, CSS, and Javascript—benefit from standardization and strong backward compatibility. Python, by contrast, is well suited for data wrangling and analysis, and has become a dominant platform in computational biology. Some libraries, such as NumPy (Harris et al. 2020) and pandas (McKinney 2010), introduced concepts now considered core to the Python ecosystem, even though they exist outside the language and its standard library. Similarly, we advocate separating the tasks of building the 3D genome models (e.g., Hi-C-based simulations) and visualizing them. This separation must be facilitated by robust mechanisms for exchanging data between tools.

Therefore, uchimata embraces existing data standards, some of which reach beyond genomics or bioinformatics. This approach allows it to benefit from mature infrastructure developed by broader communities. For example, uchimata uses Apache Arrow (https://arrow.apache.org) as its in-memory representation of 3D genome structures. Rather than supporting a fixed set of file formats, the uchimata Javascript library accepts only structures as Arrow tables, whereas the Python package also ingests canonical scientific Python structures such as NumPy array and pandas DataFrame, internally converting both to Arrow. This approach addresses incompatibilities in existing file formats for genome structures. Several tools reinterpret the PDB format, originally designed for atomistic models, to store 3D genome structures, but each adopts its own conventions. One concrete issue is that genomic coordinates can span up to nine digits, exceeding the standard column widths of the text-based PDB format and forcing users to use non-standard fields. Representing 3D genome models as Arrow tables leverages a standardized binary format widely supported by analytical tools. Our adoption of Apache Arrow is inspired by other efforts to unify bioinformatics file formats, e.g., Oxbow (Abdennur et al. 2025).

The main downside is the need to convert existing structures to Arrow. Due to the subtle differences among file formats, fully automatic conversion is rarely possible. We provide examples of how to perform this conversion in the Supplemental Materials.

## Implementation

The core web uchimata library (https://github.com/hms-dbmi/uchimata) is implemented using Typescript, transformed into a standard Javascript module, and hosted on NPM. The Jupyter notebook widget (https://github.com/hms-dbmi/uchimata-py) is built using anywidget (Manz et al. 2024) and uses the Javascript library to serve the canvas within a notebook's cell. The 3D graphics rendering is accomplished through three.js, currently using the WebGL backend. We use DuckDB (Raasveldt and Mühleisen 2019) to perform a variety of queries to filter the Apache Arrow tables representing the 3D structures and associated data.

# Results

The development of uchimata was driven by the aim to support a broad range of end-use scenarios. We focused on core functionality that could be reused in web-based applications with case-specific interfaces, without overloading the core library with features relevant only to a narrow subset of users. We also envisioned seamless integration of uchimata into data portals.

A second major group of use cases involves visualization in exploratory stages, such as during simulations and in downstream analyses of simulated data. While the library can be used directly in Javascript-based notebook environments such as Observable Notebooks (https://observablehq.com/platform/notebooks), much computational notebook work today is carried out in Python, leveraging its extensive scientific software ecosystem.

## Integration in web applications & Javascript-based notebooks

In the first scenario, uchimata functions like any other Javascript library in a web environment, enabling visualization of 3D genome structural data. To show uchimata's ability to support novel visualizations, we use it within Observable Notebooks—a literate programming environment, similar to Jupyter Notebooks, that executes Javascript code in interactive cells. A set of examples is available at https://hms-dbmi.github.io/uchimata/, source code for these examples is in the uchimata Github repository, under the `docs` folder. At its core, a 3D genome structure is a list of XYZ coordinates, sometimes associated with genomic coordinates. How these data items are visually represented is left to the user, who specifies a *view config* that maps items to concrete *marks*, such as spheres or cubes, and their visual features such as *color* and *scale*. This manner of declaring visual encodings is inspired by the grammar-of-graphics approach (Wilkinson 2005). A *scene* contains an array of structures, each paired with its corresponding view config, which together define how the structures are rendered. The view config lets users define how each structure is depicted by mapping data columns to visual attributes. For example, mapping a column containing the chromosome information to the color channel assigns a different color to each chromosome. Such annotations can be applied at any level (e.g., compartment, TADS) and depend solely on the data supplied with the structure.

We envision uchimata being integrated into use case-specific web applications and genome browsers. We recently used uchimata to extend the Gosling grammar (L'Yi et al. 2022) with support for 3D genome models (Kouřil et al. 2025), thereby implementing the previously recognized 'spatial' layout for genomic coordinate systems (Nusrat et al. 2019) that was missing in existing grammar-based tools. This grammar-of-graphics approach allows for both expressiveness and reproducibility, and by unifying 3D representations of genomes with conventional genomic data views, it opens new avenues for exploring efficient visual linking between distant genomic loci.

## Interoperability with existing bioinformatics workflows in Python

Computational notebooks play an important role in day-to-day analytical work of genomics researchers (van den Brandt et al. 2025). In the second scenario, uchimata can be used as a widget for Python-based computational notebooks.

Python is often used for chromatin simulations, where visual inspection is typically the first step in assessing the results. To shorten the iteration loop, a visualization tool should be available directly within the simulation environment, enabling biologists to perform basic checks, adjust parameters, and rerun simulations. The uchimata widget offers multiple ways to input data which aligns with in-memory storage practices in Python-based computation notebooks. Most relevant for Jupyter environments are widely used data structures such as NumPy arrays and pandas DataFrame.

Furthermore, the availability through Python allows for integration with existing bioinformatics tools. For example, we can combine uchimata with bioframe (Open2C et al. 2024) to apply genomic range selections on 3D genome structures. Bioframe offers functionality for loading genomic data from typical formats such as BED, GFF, or GTF, and performing a variety of interval operations, representing the ranges as pandas DataFrames. Uchimata then accepts these dataframes as selection queries and outputs corresponding bins of the 3D structure. The user can thus apply the same range as selection across multiple different 3D structures, facilitating comparison across an ensemble of simulated structures. In addition to genomics-based filtering, uchimata supports spatial selections (Fig. 1B), currently implementing cutting-plane (cross-section) operations and selecting spherical neighborhoods of a specified radius.

Fig. 1D highlights two examples that demonstrate how visualization of 3D models reveals insights about genome structure. First, we use Stevens et al.'s mouse cell structures (Stevens et al. 2017) and encode each chromosome's coordinates with a continuous color scale. This results in clear identification of locations where telomeres concentrate (Fig. 1D, left). Second, we aggregate gene annotation loaded from a GTF into bins matching the resolution of the structure. We then map this gene density data to both color and scale of the marks representing the bins (Fig. 1D, right). Scaling the marks can act as a form of removing occlusion and highlighting inner structure. The biological insight in this example is that gene-rich regions concentrate in the center of the genome structure, further from the border.

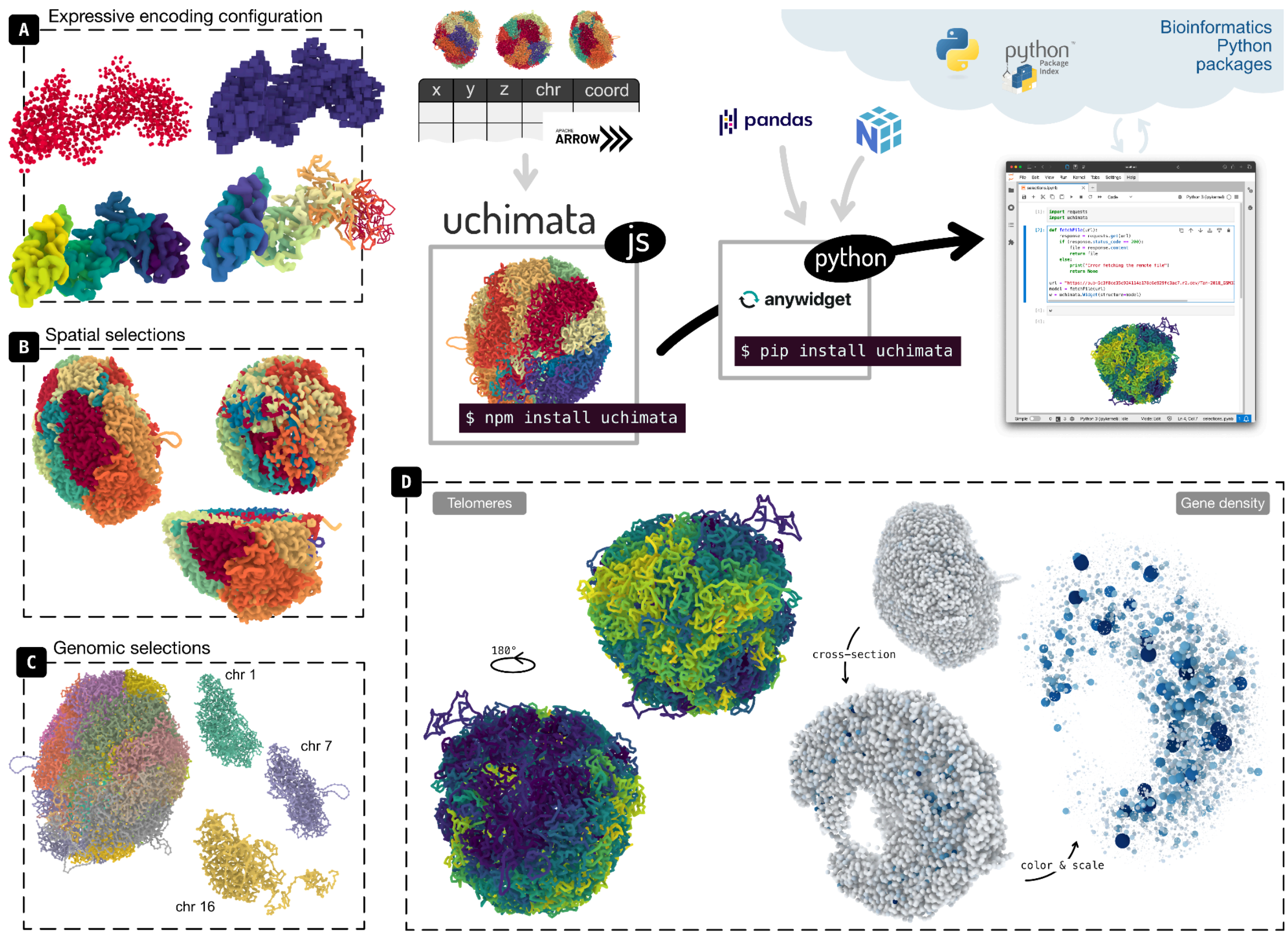

**Figure 1.** Uchimata consists of two packages, Javascript and Python, intended to cover use cases for visualization of three-dimensional genome models across the web and computational notebooks. Key features include expressive configuration of visual encoding (A), selections based on spatial attributes (B), and selections based on genomic coordinates (C). Two use cases illustrating situations where 3D visualization brings new insights (D): mapping bin coordinates to continuous color scale shows concentration of telomeres on one side of the structures (left), and encoding gene density to color and scale of the bin marks shows concentration of gene-rich regions within the structures, further from the boundary (right).

# Conclusion

While the heatmap representation (i.e., the contact matrices resulting from Hi-C) remains the most direct experimental way to observe how whole genomes fold in nuclear space, we believe that the three-dimensional structural representation offers a complementary benefit. Its primary strength is that it contextualizes genomic data within 3D space, which can lead to hypotheses related to specific patterns observed in the actual 3D structure. With uchimata, we contribute software that makes it easier to visualize this type of genomics data. With this crucial infrastructure in place, we intend to further investigate means of linking between traditional genome browser views, dense multiscale matrix viewers, and three-dimensional structure visualizations. On the rendering side, we plan to adopt the WebGPU API as it becomes broadly supported across major web browsers.

## Acknowledgements

This work was supported in part by the National Institutes of Health (R01 HG011773 and UM1 HG011536).

## References

Abdennur, Nezar, conradbzura, Trevor Manz, et al. 2025. *Abdenlab/oxbow: Py-oxbow@v0.5.0*. Version py-oxbow@v0.5.0. Zenodo. https://doi.org/10.5281/ZENODO.17642721.

Asbury, Thomas M., Matt Mitman, Jijun Tang, and W. Jim Zheng. 2010. “Genome3D: A Viewer-Model Framework for Integrating and Visualizing Multi-Scale Epigenomic Information within a Three-Dimensional Genome.” *BMC Bioinformatics* 11 (1): 444.

Berman, H. M., J. Westbrook, Z. Feng, et al. 2000. “The Protein Data Bank.” *Nucleic Acids Research* 28 (1): 235–242.

Brandt, Astrid van den, Sehi L’Yi, Huyen N. Nguyen, Anna Vilanova, and Nils Gehlenborg. 2025. “Understanding Visualization Authoring Techniques for Genomics Data in the Context of Personas and Tasks.” *IEEE Transactions on Visualization and Computer Graphics* 31 (1): 1180–1190.

Butyaev, Alexander, Ruslan Mavlyutov, Mathieu Blanchette, Philippe Cudré-Mauroux, and Jérôme Waldispühl. 2015. “A Low-Latency, Big Database System and Browser for Storage, Querying and Visualization of 3D Genomic Data.” *Nucleic Acids Research* 43 (16): e103.

Djekidel, Mohamed Nadhir, Mengjie Wang, Michael Q. Zhang, and Juntao Gao. 2017. “HiC-3DViewer: A New Tool to Visualize Hi-C Data in 3D Space.” *Quantitative Biology (Beijing, China)* 5 (2): 183–190.

Duan, Zhijun, Mirela Andronescu, Kevin Schutz, et al. 2010. “A Three-Dimensional Model of the Yeast Genome.” *Nature* 465 (7296): 363–367.

Harris, Charles R., K. Jarrod Millman, Stéfan J. van der Walt, et al. 2020. “Array Programming with NumPy.” *Nature* 585 (7825): 357–362.

Hunter, John D. 2007. “Matplotlib: A 2D Graphics Environment.” *Computing in Science & Engineering* 9 (3): 90–95.

Imakaev, Maxim V., Geoffrey Fudenberg, and Leonid A. Mirny. 2015. “Modeling Chromosomes: Beyond Pretty Pictures.” *FEBS Letters* 589 (20 Pt A): 3031–3036.

Ko, Amy J., Robin Abraham, Laura Beckwith, et al. 2011. “The State of the Art in End-User Software Engineering.” *ACM Computing Surveys* 43 (3): 1–44.

Kouřil, David, Trevor Manz, Sehi L’Yi, and Nils Gehlenborg. 2025. “Design Space and Declarative Grammar for 3D Genomic Data Visualization.” May 8. https://doi.org/10.31219/osf.io/dtr6u_v1.

Li, Daofeng, Deepak Purushotham, Jessica K. Harrison, et al. 2022. “WashU Epigenome Browser Update 2022.” *Nucleic Acids Research* 50 (W1): W774–W781.

Lieberman-Aiden, Erez, Nynke L. van Berkum, Louise Williams, et al. 2009. “Comprehensive

Mapping of Long-Range Interactions Reveals Folding Principles of the Human Genome.” *Science (New York, N.Y.)* 326 (5950): 289–293.

Li, Zilong, Stephanie Portillo-Ledesma, and Tamar Schlick. 2023. “Techniques for and Challenges in Reconstructing 3D Genome Structures from 2D Chromosome Conformation Capture Data.” *Current Opinion in Cell Biology* 83 (102209): 102209.

L’Yi, Sehi, Qianwen Wang, Fritz Lekschas, and Nils Gehlenborg. 2022. “Gosling: A Grammar-Based Toolkit for Scalable and Interactive Genomics Data Visualization.” *IEEE Transactions on Visualization and Computer Graphics* 28 (1): 140–150.

Manz, Trevor. 2024. “Composable Visualization Systems for Biological Data.” Harvard University.

Manz, Trevor, Nezar Abdennur, and Nils Gehlenborg. 2024. “Anywidget: Reusable Widgets for Interactive Analysis and Visualization in Computational Notebooks.” *Journal of Open Source Software* 9 (102): 6939.

McKinney, Wes. 2010. “Data Structures for Statistical Computing in Python.” *Proceedings of the Python in Science Conference*, 56–61.

Nguyen, Hai, David A. Case, and Alexander S. Rose. 2018. “NGLview-Interactive Molecular Graphics for Jupyter Notebooks.” *Bioinformatics (Oxford, England)* 34 (7): 1241–1242.

Nowotny, Jackson, Avery Wells, Oluwatosin Oluwadare, et al. 2016. “GMOL: An Interactive Tool for 3D Genome Structure Visualization.” *Scientific Reports* 6 (1): 20802.

Nusrat, S., T. Harbig, and N. Gehlenborg. 2019. “Tasks, Techniques, and Tools for Genomic Data Visualization.” *Computer Graphics Forum: Journal of the European Association for Computer Graphics* 38 (3): 781–805.

Oluwadare, Oluwatosin, Max Highsmith, and Jianlin Cheng. 2019. “An Overview of Methods for Reconstructing 3-D Chromosome and Genome Structures from Hi-C Data.” *Biological Procedures Online* 21 (1): 7.

Oluwadare, Oluwatosin, Max Highsmith, Douglass Turner, Erez Lieberman Aiden, and Jianlin Cheng. 2020. “GSDB: A Database of 3D Chromosome and Genome Structures Reconstructed from Hi-C Data.” *BMC Molecular and Cell Biology* 21 (1): 60.

Open2C, Nezar Abdennur, Geoffrey Fudenberg, et al. 2024. “Bioframe: Operations on Genomic Intervals in Pandas Dataframes.” *Bioinformatics (Oxford, England)* 40 (2). https://doi.org/10.1093/bioinformatics/btae088.

Portillo-Ledesma, Stephanie, Zilong Li, and Tamar Schlick. 2023. “Genome Modeling: From Chromatin Fibers to Genes.” *Current Opinion in Structural Biology* 78 (102506): 102506.

Raasveldt, Mark, and Hannes Mühleisen. 2019. “DuckDB.” Paper presented SIGMOD/PODS ’19: International Conference on Management of Data, Amsterdam Netherlands. *Proceedings of the 2019 International Conference on Management of Data* (New York, NY, USA), June 25. https://doi.org/10.1145/3299869.3320212.

Schrödinger, LLC. 2015. “The PyMOL Molecular Graphics System, Version 1.8.” Preprint, November.

Schuette, Greg, Zhuohan Lao, and Bin Zhang. 2025. “ChromoGen: Diffusion Model Predicts Single-Cell Chromatin Conformations.” *Science Advances* 11 (5): eadr8265.

Sehnal, David, Sebastian Bittrich, Mandar Deshpande, et al. 2021. “Mol* Viewer: Modern Web App for 3D Visualization and Analysis of Large Biomolecular Structures.” *Nucleic Acids Research* 49 (W1): W431–W437.

Stevens, Tim J., David Lando, Srinjan Basu, et al. 2017. “3D Structures of Individual Mammalian Genomes Studied by Single-Cell Hi-C.” *Nature* 544 (7648): 59–64.

Tan, Longzhi, Dong Xing, Chi-Han Chang, Heng Li, and X. Sunney Xie. 2018. “Three-Dimensional Genome Structures of Single Diploid Human Cells.” *Science (New York, N.Y.)* 361 (6405): 924–928.

Todd, Stephen, Peter Todd, Simon J. McGowan, et al. 2021. “CSynth: An Interactive Modelling and Visualization Tool for 3D Chromatin Structure.” *Bioinformatics (Oxford, England)* 37 (7): 951–955.

Wilkinson, Leland. 2005. *The Grammar of Graphics*. 2nd ed. Statistics and Computing. Springer. PDF.

Zhu, Xiaopeng, Yang Zhang, Yuchuan Wang, et al. 2022. “Nucleome Browser: An Integrative and Multimodal Data Navigation Platform for 4D Nucleome.” *Nature Methods* 19 (8): 911–913.